\documentclass[conference]{IEEEtran}

\pdfoutput=1

\usepackage{subfigure}
\usepackage{color}
\usepackage{graphicx}
\usepackage{amssymb}

\def\QED{\mbox{\rule[0pt]{1.5ex}{1.5ex}}} 
\def\endproof{\hspace*{\fill}~\QED\par\endtrivlist\unskip} 
\newtheorem{definition}{Definition} 
\newtheorem{lemma}{Lemma} 
\newtheorem{property}{Property} 
\newtheorem{theorem}{Theorem} 
\newtheorem{equivalence}{Equivalence} 
\newtheorem{constraint}{Constraint} 
\newtheorem{remark}{Remark} 

\begin{document}

\title{Node Coloring in Wireless Networks:\\ Complexity Results and Grid Coloring}
\author{\IEEEauthorblockN{Ichrak Amdouni, Pascale Minet and Cedric Adjih}
\IEEEauthorblockA{INRIA\\
Rocquencourt\\
78153 Le Chesnay cedex, France\\
Email: ichrak.amdouni@inria.fr, pascale.minet@inria.fr, cedric.adjih@inria.fr}}

\maketitle

\pagestyle{plain} 

\begin{abstract}
Coloring is used in wireless networks to improve communication efficiency, mainly in terms of bandwidth, energy and possibly end-to-end delays. In this paper, we define the $h$-hop node coloring problem, with $h$ any positive integer, adapted to two types of applications in wireless networks. We specify both general mode for general applications and strategic mode for data gathering applications. We prove that the associated decision problem is NP-complete. We then focus on grid topologies that constitute regular topologies for large or dense wireless networks. We consider various transmission ranges and identify a color pattern that can be reproduced to color the whole grid with the optimal number of colors.  We obtain an optimal periodic coloring of the grid for the considered transmission range. We then present a 3-hop distributed coloring algorithm, called SERENA. Through simulation results, we highlight the impact of node priority assignment on the number of colors obtained for any network and grids in particular. We then compare these optimal results on grids with those obtained by SERENA and identify directions to improve SERENA.
\end{abstract}

\section{Motivations\label{Motivations}}
Coloring has been used in wireless ad hoc and sensor networks to improve communications efficiency by scheduling medium access. Indeed, only nodes that do not interfere can have the same color and then are allowed to transmit simultaneously. Hence coloring can be used to schedule node activity. The expected benefits of coloring are threefold:
\begin{enumerate}
\item At the bandwidth level where no bandwidth is lost in collisions, overhearing and interferences are reduced. Moreover, the use of the same color by several nodes ensures the spatial reuse of the bandwidth \cite{rajendran03, rajendran05, TDMA-ASAP, serena, FlexiTP}.
\item At the energy level where no energy wasted in collision. Furthermore, nodes can sleep to save energy without loosing messages sent to them because of the node activity schedule based on colors \cite{rajendran03, rajendran05, TDMA-ASAP, serena, ZMAC, FlexiTP}.
\item At the delay level where the end-to-end delays can be optimized by a smart coloring ensuring for instance that any child accesses the medium before its parent in the data gathering tree \cite{rajendran05, TDMA-ASAP, serena, FlexiTP}.
\end{enumerate}

Concerning coloring algorithms, two types of coloring are distinguished:
node coloring and link coloring. With link coloring, time slots are assigned per link. Only the transmitter and the receiver are awake, the other nodes can sleep. If the link is lightly loaded, its slot can be underused. Moreover, broadcast communications are not easy: the source must send a copy to each neighbor. On the contrary, with node coloring, the slot is assigned to the transmitter that can use it according to its needs: unicast and/or broadcast transmissions. Hence the slot use is optimized by its owner.

The value of $h$ in $h$-hop node coloring must be chosen to ensure that any two nodes that are strictly more than $h$-hop away can transmit simultaneously without interfering. It follows that $h$ depends on the type of communication that must be supported. For instance, broadcast transmissions require 2-hop coloring, whereas unicast transmission with immediate acknowledgement (i.e. the receiver uses the timeslot of the sender to transmit its acknowledgement) requires 3-hop coloring, see \cite{serena} for an illustrative example.

This paper is organized as follows. First, we define the coloring problem in Section~\ref{Problem} and introduce definitions. We position our work with regard to the state of the art in Section~\ref{StateArt}. In Section~\ref{Complexity}, we prove that the $h$-hop coloring decision problem is NP-complete in both general and strategic modes, for any integer $h>0$. That is why, we propose SERENA, an heuristic to color network nodes.  In Section~\ref{Theoretical}, we obtain theoretical results determining an optimal periodic color pattern for grid topologies with various transmission ranges. We compare them with SERENA results in Section~\ref{SERENA}. Finally, we conclude in Section~\ref{Conclusion} pointing out future research directions.

\section {Coloring problem definition \label{Problem}}
\subsection{Generalities \label{general}}
Let $G(V,E)$ be an undirected graph representing the network topology. Each vertex $v_i \in V$ represents a network node with $i \in [1,n]$, where $n$ is the number of network nodes. For all vertices $v_1$ and $v_2$ in $V$, the edge $(v_1,v_2) \in E$ if and only if the two nodes $v_1$ and $v_2$ are one-hop neighbors. There are two types of graph coloring:
vertex (or node) coloring assigns a color to each vertex of the graph, whereas edge (or link) coloring assigns a color to each edge of the graph.
\begin{definition}
\textbf{One-hop node coloring} of $G$ consists
in (i) assigning each vertex in $V$ a color in such a way that two adjacent vertices have different colors and (ii) using the smallest number of colors.
\end{definition}


We can easily extend one-hop node coloring to $h$-hop node coloring, where $h$ is an integer strictly positive. $h$-hop node coloring is also called $h$-distance node coloring and 1-hop coloring of the power $h$ of the graph $G$ in \cite{kdist}. 
\begin{definition}
For any integer $h>0$, two nodes $u$ and $v$ are said $h$-hop neighbors if and only if the shortest path between them has a length of $h$ links.
\end{definition}
\begin{definition}
For any integer $h>0$, a $h$-hop node coloring is said \textbf{valid} if and only if any two nodes that are $k$-hop neighbors, with $1 \leq k\leq h$ have not the same color.
\end{definition}
\begin{definition}
A valid $h$-hop node coloring is said \textbf{optimal} if and only if no valid $h$-hop node coloring uses less colors that this coloring.
\end{definition}
It can be easily shown that the breakage of a link does not compromise the validity of a coloring, whereas the creation of a new link may.\\
For simplicity reasons, colors are represented by natural integers, starting with zero.

\subsection{Application to wireless networks}

\begin{figure}[!h]
	\begin{center}
	\includegraphics[width=0.3\linewidth]{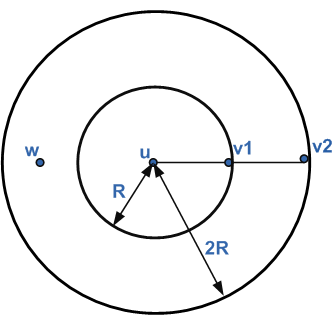}
		\vspace{-7pt}
	\caption{Example of 1-hop and 2-hop nodes of $u$.} \label{Fig:1or2hopnodes}
	\end{center}
\end{figure}

We can notice that almost all coloring algorithms when applied to a wireless network, make the assumption of an ideal wireless environment.
\begin{definition}
A wireless environment is said \textbf{ideal} if and only if:
\begin{itemize}
\vspace{-3pt}\item Any node has a unique address in the wireless network.
\item Links are symmetric: if node $u$ sees node $v$ as a one-hop neighbor, then conversely node $v$ sees node $u$ as a one-hop neighbor.
\item Links are stable. More precisely, link creation during or after the completion of the coloring algorithm is not taken into account.
\item For any node $u$, any node $w$ that is not in transmission range of $u$ cannot prevent $u$ from correctly receiving a message sent by $v$ of its one-hop neighbors.
\end{itemize}
\end{definition}
\begin{definition}
For any node $u$, its 1-hop neighbors are all the nodes located in the disk centered at node $u$ of radius the transmission range $R$.
For any integer $h>1$, any two nodes $u$ and $v$ are $h$-hop neighbors if and only if $u$ is $(h-1)$-hop away from a 1-hop node of $v$.
\end{definition}
Figure~\ref{Fig:1or2hopnodes} depicts a node $w$ that belongs to the disk centered at node $u$ of radius $2R$. Node $w$ is not a $2$-hop node of $u$, whereas $v_2$ is because of the existence of $v_1$. Consequently, with a 2-hop coloring, $v_1$ and $v_2$ cannot have the same color as $u$, whereas $w$ can. If now we consider a transmission range $R'=2R$ and a 1-hop coloring, nodes $v_1$, $v_2$ and $w$ cannot have the same color as $u$.

\vspace*{-10pt}

\begin{figure}[!h]
	\begin{center}
	\subfigure[]{\includegraphics[width=0.32\linewidth]{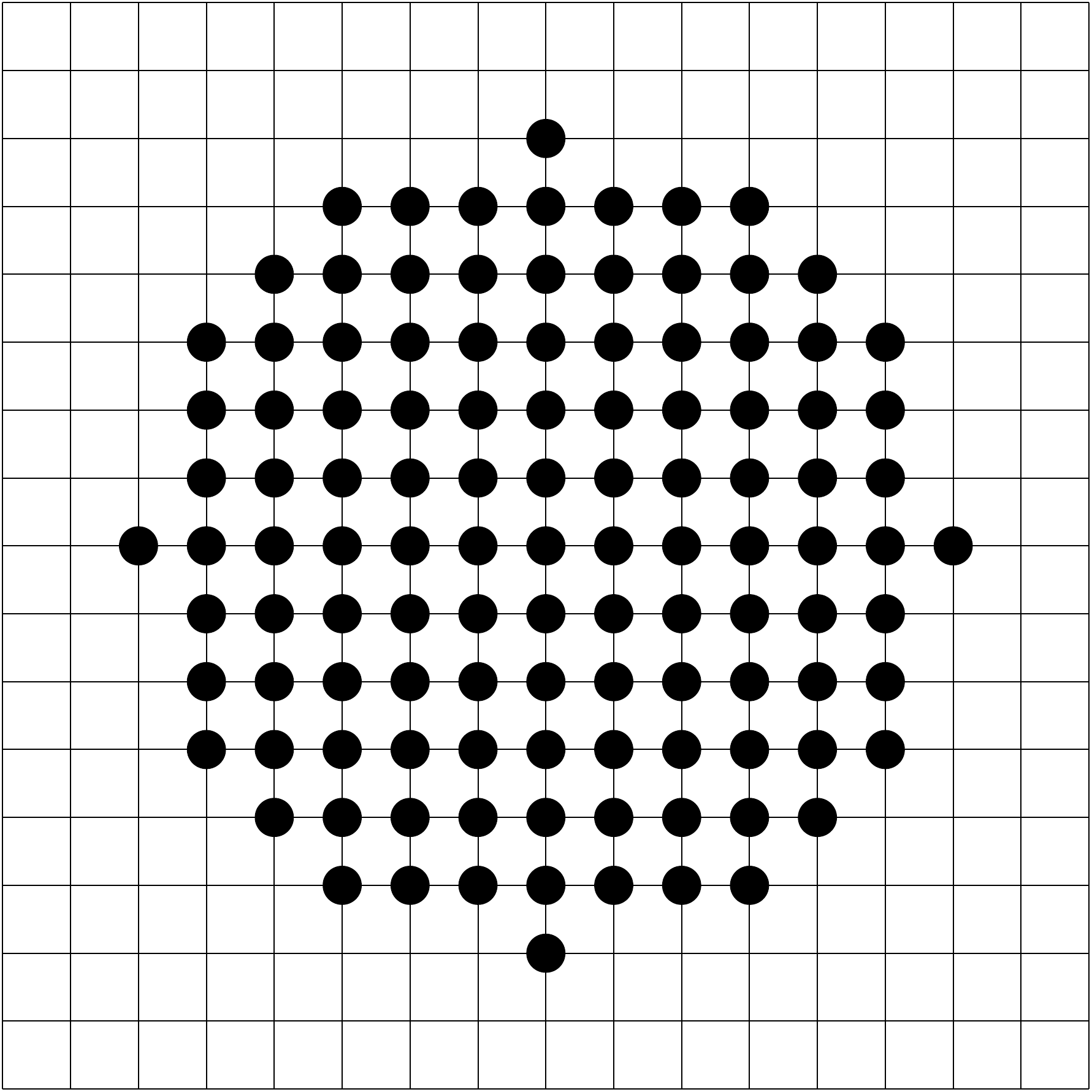}\label{H1R6}}
	\subfigure[]{\includegraphics[width=0.32\linewidth]{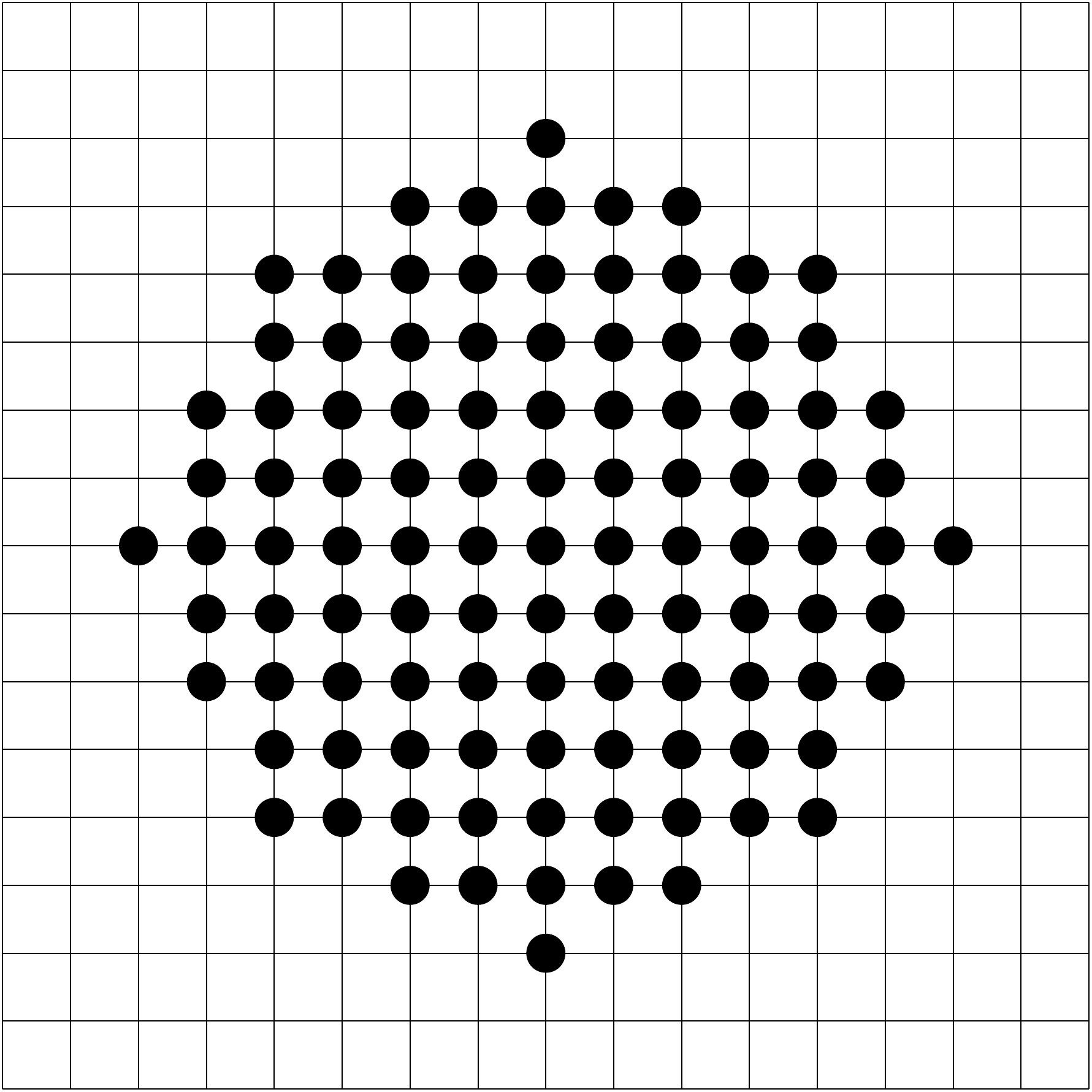}\label{H2R3}}
	\subfigure[]{\includegraphics[width=0.32\linewidth]{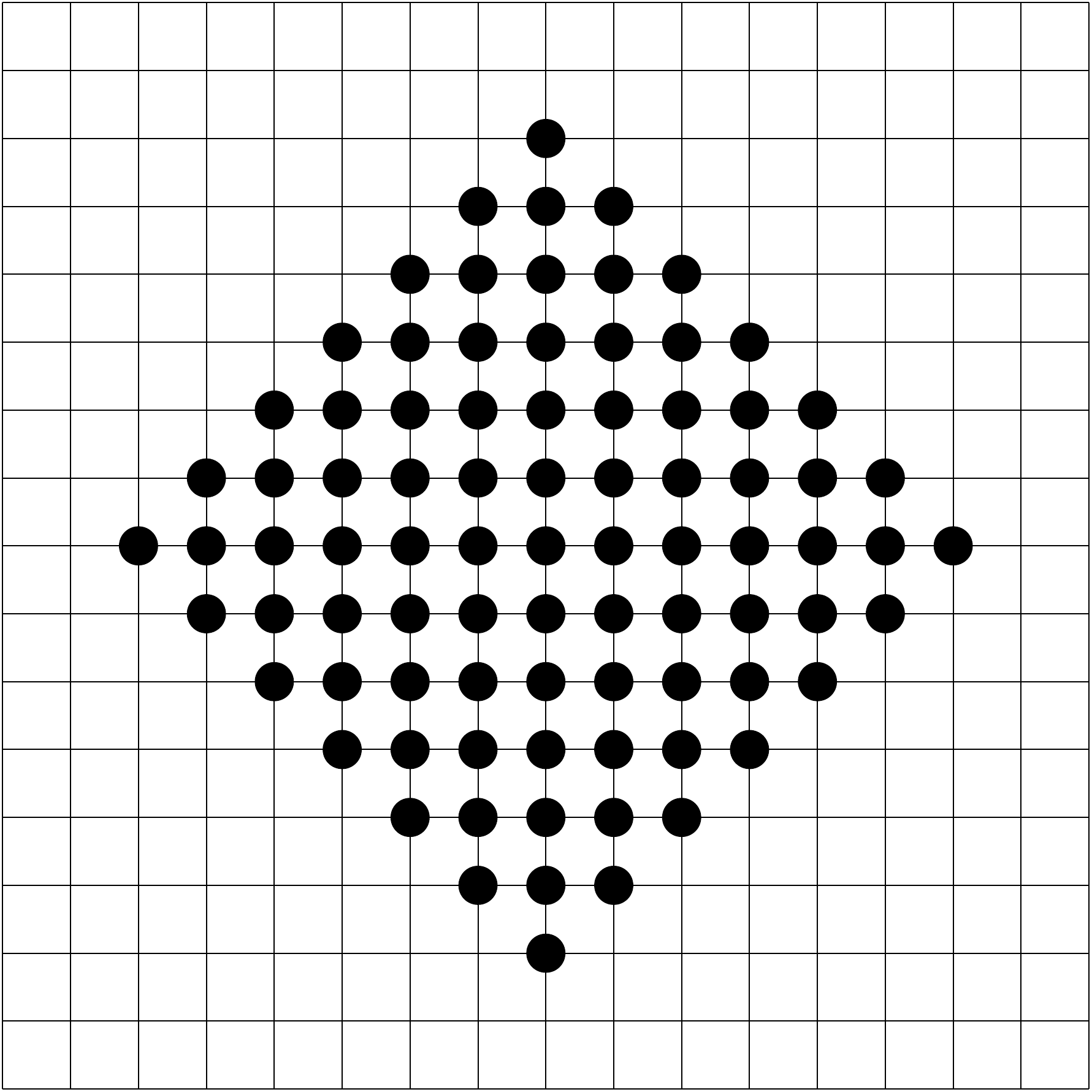}\label{H3R2}}
	\vspace{-7pt}
	\caption{Nodes having a color different from $u$: (a) 1-hop coloring $R=6$; (b) 2-hop coloring $R=3$ and (c) 3-hop coloring $R=2$.}     \label{Fig:Rxh}
	\end{center}
\end{figure}

\vspace*{-20pt}
The following example is taken in grid topologies, where the transmission range is expressed in grid step units.  For any grid node $u$, Figure~\ref{Fig:Rxh} depicts the nodes that cannot take the same color as $u$ in case a) of a 1-hop coloring with transmission range=6, in case b) of 2-hop coloring with transmission range=3 and finally in case c) of 3-hop coloring with transmission range=2.

\vspace*{-8pt}
	
\vspace*{-12pt}
\begin{remark}
\label{remark1}
For any graph $G$, for any transmission ranges $R\ and\ R' \geq 1$, for any integer $h\ and\ k >0$ such that $R' \cdot k = R \cdot h$, $h$-hop coloring of $G$ with transmission range $R$ is not equivalent to $k$-hop coloring of $G$ with transmission range $R'$.
\end{remark}

To be able to support different applications, two modes are introduced in $h$-hop node coloring:
\begin{itemize}
\item \textit{the general mode} to support general applications, where the destination of any unicast transmission can be any network node.
The general mode is the default mode of the $h$-hop node coloring problem defined in section~\ref{general}.
\item \textit{the strategic mode} to support data gathering applications, where the destination of any unicast transmission is either the parent or a child of the node in the data gathering tree. With regard to the general mode, the strategic mode requires an additional constraint to be met:
\begin{constraint}
\label{C0}
Any node must select a color strictly higher than the color taken by its parent in the data gathering tree.
\end{constraint}
\end{itemize}
To compare the performance of coloring algorithms, two criteria are used:
\begin{itemize}
\item the number of colors used: the optimal number is called chromatic number.
\item the number of rounds needed to color the graph. By definition, in a round, a node is able to send a message to its neighbors, to receive their message and to process them.
\end{itemize}
We now present a brief state of the art dealing with graph coloring and its application to radio networks and wireless sensor networks.

\section{State of the art\label{StateArt}}
Coloring has been first introduced in graphs with vertex coloring and edge coloring.
One-hop vertex coloring has been shown NP-complete in \cite{gar79} for the general case, whereas graphs with maximum vertex degree less than four and bipartite graphs can be colored in polynomial time. The first algorithms proposed were centralized like the greedy algorithm Dsatur (no color backtracking)~\cite{brel79}, where the vertex with the highest number of already colored neighbor vertices is colored first. Later on,
decentralized ones like Distributed Largest First \cite{hans04} were designed. In this algorithm, each node selects a color. It is allowed to keep its color only if it has the largest degree among its neighbors but also if this color does not conflict with the colors of its neighbors. This algorithm runs in O($\Delta^2 log n$), where $\Delta$ is the largest vertex degree and $n$ the number of vertices. The algorithm given in \cite{kuhn06} proceeds iteratively by reducing the number of colors step-by-step; initially, all nodes have distinct colors. This algorithm runs in O($\Delta^2 log n$) and uses a number of colors close to $\Delta$.\\

\vspace{-8pt}The theoretical performance of coloring algorithms has been studied in the
literature, in general for 2-hop coloring.
Because minimum 2-hop coloring is NP-complete, the focus has been on evaluating
the performance of \emph{approximation} algorithms, whose objective
is not to find the optimal coloring, but at least to be reasonably close.
A typical approximation algorithm for coloring is \emph{FirstFit}:
\emph{FirstFit} \cite{Car07} sequentially assigns colors to nodes; it chooses for each
node the first available color.
Depending of the order in which the nodes are colored,
different results (with varied performance) are obtained.\\

\vspace{-8pt}Edge coloring problems can be transformed into a vertex version: an edge coloring of a graph is just a vertex coloring of its link graph. Applied to wireless networks, edge coloring has been called link scheduling. For instance,~\cite{Gandham08} obtains a TDMA MAC schedule enabling two-way communication between every pair of neighbors. Edges are colored and a 
feasible direction of transmission is found.\\

\vspace{-8pt}Coloring has then been applied to radio networks to provide a collision-free medium access. The goal is to schedule transmissions in time slots, in such a way that two senders allowed to transmit in the same slot do not interfere. Such a problem is also called broadcast scheduling in~\cite{Rama89}, channel assignment in ~\cite{Krumke00} or slot assignment in ~\cite{interfcolor}. TRAMA,~\cite{rajendran03}, schedules node transmissions by assigning time slots to transmitting nodes and receivers. Only nodes having data to send contend for a slot. The node with the highest priority in its neighborhood up to 2-hop wins the right to transmit in the slot considered. Each node declares in advance its next schedule containing the list of its slots and for each slot its receiver(s). The adaptivity of TRAMA to the traffic rate comes at a price: its complexity. DRAND,~\cite{DRAND}, the coloring algorithm used with the hybrid ZMAC protocol, \cite{ZMAC}, that operates like CSMA under low contention and like TDMA otherwise, assigns slots to nodes in such a way that 1-hop and 2-hop neighbors have different slots. This randomized algorithm has the advantage of not depending on the number of nodes but at the cost of an asymptotic convergence.\\

\vspace{-8pt}More recently, coloring algorithms have been designed for WSNs, like FLAMA~\cite{rajendran05}, an adaptation of TRAMA, where its overhead has been considerably reduced. This is obtained by supporting communications limited to the data gathering tree, like TDMA-ASAP. TDMA-ASAP \cite{TDMA-ASAP} integrates a coloring algorithm with the medium access. Moreover, this protocol can adapt to various traffic conditions by allowing a node to steal an unused slot to its brother in the tree. In WSNs where energy matters, sensors receive
consecutive time slots, like~\cite{Ma09}, to reduce the frequency of state
transitions and thus save energy. FlexiTP~\cite{FlexiTP} is a TDMA-based protocol in
which a slot is assigned to one transmitter and one receiver. 
The slot assignment order is given by a deep-first search of the tree.
A node selects the smallest available slot in its neighborhood up to 2 hops and advertises its schedule.
 This solution does not support immediate acknowledgment.
In \cite{serena}, we proposed SERENA a node coloring algorithm that increases energy efficiency by avoiding collisions, reducing overhearing, allowing nodes to sleep to save energy and enabling spatial reuse of the bandwidth. This algorithm can support various types of communication (unicast with immediate acknowledgement, broadcast). It can also be optimized for data gathering applications by scheduling the children before their parent. Notice that SERENA is a practical distributed protocol that implements
an efficient  version of the algorithm \emph{FirstFit} for a $3$-hop coloring,
with a specific order induced by node priority.
In this paper, we give new theoretical results for grid coloring. Finally, we show how SERENA can be optimized.\\


\vspace{-10pt}\section{Complexity study \label{Complexity}}
In this section, we will demonstrate that $h$-hop ($h\ge1$) vertex coloring in both general and strategic modes is a NP-complete problem. This assertion is given by Theorem~\ref{Thcomplexity}:
\begin{theorem}
\label{Thcomplexity}
The decision problem of $h$-hop (h$\ge$1) vertex coloring in both general and strategic modes is NP-complete.\\
\end{theorem}

\vspace*{-8pt}
It has been proved in~\cite{gar79} that the 1-hop vertex coloring problem is NP-complete. We now prove the NP-completeness in both general and strategic modes for $h \geq 1$. Our proof of Theorem~\ref{Thcomplexity} is based on the following steps:\\

\vspace*{-8pt}
\noindent$\bullet$ First, we define the associated decision problem of the $h$-hop vertex coloring of a graph $G$ in both general and strategic modes: can graph $G$ be colored with $k$ colors in both general and strategic modes ($k$ is a positive integer smaller than the vertex number), such that two nodes that are $l$-hop neighbors with $1 \le l \le h$ have not the same color? This problem is called \textbf{$k$-color $h$-hop coloring in both general and strategic modes}.\\

\vspace*{-8pt}
\noindent$\bullet$ Second, we prove the following lemma:
\begin{lemma}
The $k$-color $h$-hop coloring problem in both general and strategic modes for $h \geq1$ is in NP.
\end{lemma}
\IEEEproof Given a $h$-hop coloring of $G$, $h \ge 1$ we can check in polynomial time
($O(n^h)$, where $n$ is the number of nodes) that the coloring
produced by a given $h$-hop algorithm in both general and strategic modes does not assign the same
color to two nodes that are $p$-hop neighbors with $1 \le p \le h$, and that the
total number of colors is $k$.
\endproof

\vspace*{5pt}\noindent$\bullet$ Third, we define a reduction $f$ of the $k$-color 1-hop vertex coloring problem that has been shown NP-complete in~\cite{gar79}, to a $k'$-color $h$-hop coloring problem in both general and strategic modes, with $k'$ a positive integer smaller than the nodes number. This reduction should be polynomial in time. Based on this reduction, we then prove the equivalence:
\begin{equivalence}
\label{equiv}
A $k'$-color $h$-hop vertex coloring problem in both general and strategic modes has a solution if and only if a $k$-color 1-hop vertex coloring problem has a solution.\\
\end{equivalence}

\vspace*{-8pt}
In general, to demonstrate that a problem is NP-complete based on another problem that is known to be NP-complete, the required reduction should allow us to show that we can find a solution for the first problem if and only if we can find a solution for the second problem.
In our case, we should transform a graph $G(V,E)$ into a graph $G'=(V',E')$ and show that finding a $k$-color 1-hop coloring of $G(V,E)$ can lead to find a $k'$-color $h$-hop coloring of $G'(V',E')$ in both general and strategic modes and vice versa, proving Equivalence~\ref{equiv}.\\
Finding a valid $k'$-color $h$-hop coloring of $G'(V',E')$ in both general and strategic modes based on a valid $k$-color 1-hop coloring of $G(V,E)$ requires that for any two nodes $v_1$ and $v_2$ in $G$, the following constraints are met:
\begin{constraint}
\label{C1}
Any two nodes $v_1$ and $v_2$, 1-hop away in $G$ must be at most $h$-hop away in $G'$.
\end{constraint}
Thus, these two nodes that are assigned different colors by a 1-hop coloring of $G$ are also assigned different colors by a $h$-hop coloring of $G'$.
\begin{constraint}
\label{C2}
Similarly, any two nodes $v_1$ and $v_2$, 2-hop away in $G$ must be at least ($h+1$)-hop away in $G'$.
\end{constraint}

Consequently, the reduction separates any two 1-hop neighbors $v_i$ and $v_j$ of the initial graph $G$ by a set of nodes such that the distance between them in the new graph $G'$ is at most $h$ hops. $V'$ is obtained from $V$ by adding new nodes. We denote $V' = V \cup V_1$. The definition of these new nodes depends on $h$ parity.\\
In order to simplify the determination of $k'$, the number of colors used for the $h$-hop coloring of $G'$, we add to the transformation a new constraint:
\begin{constraint}
\label{C3}
Any two nodes in $V_1$ must be at most $h$-hop away. Moreover, any two nodes $u \in V$ and $v \in V_1$ must be at most $h$-hop away.
\end{constraint}
Thus, in a $h$-hop coloring of $G'$, nodes in $V_1$ cannot reuse a color already used in $V_1$. Similarly, no node in $V$ can reuse a color used by a node in $V_1$.\\

\begin{figure}[!h]
	\begin{center}
	\subfigure[]{\includegraphics[width=0.65\linewidth]{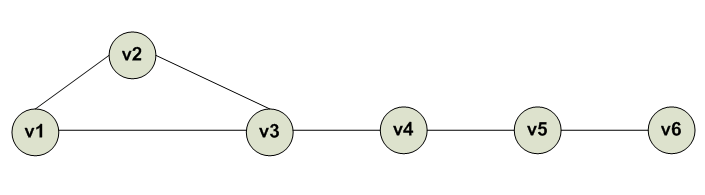}\label{originalGraph}}
	\subfigure[]{\includegraphics[width=0.9\linewidth]{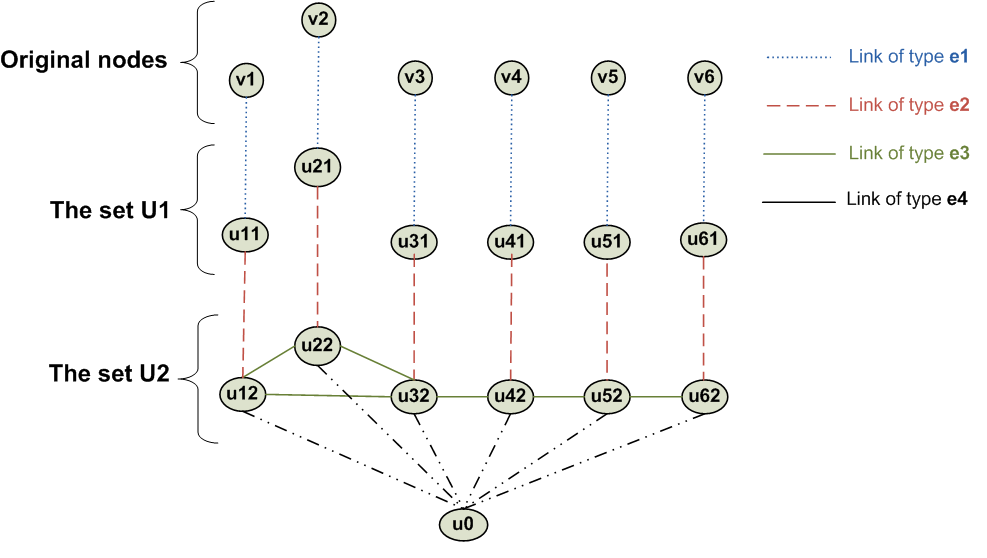}\label{5hop}}
	\vspace{-7pt}
	\caption{Example of: (a) Graph $G$; (b) Transformed graph $G'$ for $h=5$.} \label{5hopGraph}
	\end{center}
	\end{figure}

\noindent The transformation proceeds as follows, depending on the parity of $h$:\\
\textbf{- First case: $h$ is odd}: see the example of $h=5$ in Figure~\ref{5hopGraph}.\\
\indent $\bullet$ \textit{Definition of $V'$}\\
In this case, we first define $h'=(h-1)/2$ copies of $U$, denoted $U_i$ and $h'$ bijective functions $f_i$, with $i \in[1,h']$:\\
		$\begin{array}{ccccc}
		f_i & : & V & \to & U_i\\
		& & v & \mapsto & f_i(v) = u_i\\
		\end{array}$\\
Now, we can define the set $V_1 = \cup_i U_i\ \cup \{u_0\}$ and $V' = V \cup V_1$, $\forall i \in [1,h']$, where $u_0$ is a new node introduced to meet constraint C3. Node $u_0$ is a neighbor of all nodes in $U_{h'}$.\\

$\bullet$ \textit{Definition of $E'$}\\
To build the set $E'$, four types of links are introduced. We then have:  $E'= E_1 \cup E_2 \cup E_3 \cup E_4$ where:
\begin{itemize}
\item $E_1 = \{(v,u_1)\ such\ that\ v \in V \ and\ u_1=f_1(v) \in U_1\}$. Thus, each node $v$ from the initial graph $G$ is linked to $u_1$, its associated node from the set $U_1$ (see links of type $e_1$ in Figure~\ref{5hopGraph}).
\item $E_2 = \cup_{l \in[1,h'-1]} \{(u_l,u_{l+1})\ such\ that\ u_l \in U_l\ and\ u_{l+1} \in U_{l+1}\ and\ f_l^{-1}(u_l)=f_{l+1}^{-1}(u_{l+1})\}$. Each node $u_j$ from $U_j$ is linked to node $u_{j+1}$ from $U_{j+1}$ associated with the same node $v \in V$, that is $f_{j+1}^{-1}(u_{j+1})=v\ and\ f_j^{-1}(u_{j})=v$ (see links of type $e_2$).
\item $E_3 = \{(u_{h'},v_{h'})\ such\ that\ u_{h'}\ and\ v_{h'}\in U_{h'}\ and\ (f_{h'}^{-1}(u_{h'}), f_{h'}^{-1}(v_{h'})) \in E\}$. Two nodes $u_{h'}$ and $v_{h'}$ from $U_{h'}$ are linked to each other if their corresponding nodes in $V$ are linked in $E$ (see links of type $e_3$).
\item $E_4 = \{(u,u_0)\ with\ u \in U_{h'}\}$. Finally, the nodes in $U_{h'}$ are linked to the conjunction node $u_0$, which was added to meet Constraint~\ref{C3} (see links of type $e_4$).
\end{itemize}
This construction is polynomial in time. An example of graphs $G$
	and $G'$ with $h=5$ is illustrated in Figure~\ref{5hopGraph}.
	
	\vspace*{-12pt}
	
\begin{figure}[!h]
	\begin{center}
	\includegraphics[width=0.9\linewidth]{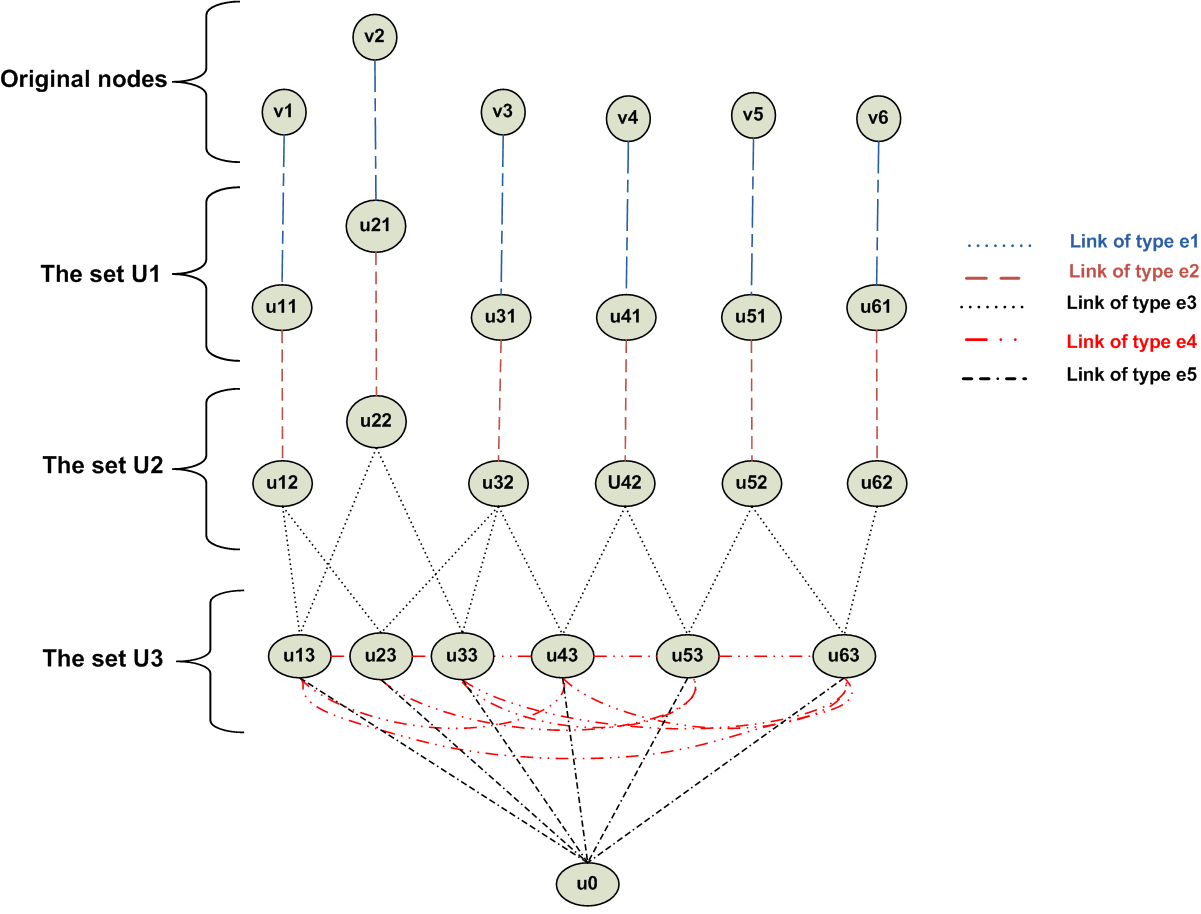}\label{6hop}
	\vspace{-7pt}
	\caption{Transformed graph $G'$ of $G$ for $h=6$.} \label{6hopGraph}
	\end{center}
\end{figure}

\vspace*{-7pt}
\textbf{- Second case: $h$ is even}: see the example of $h=6$ in Figure~\ref{6hopGraph}.\\
To build the graph $G'$ when $h$ is even, Constraints~\ref{C1}, \ref{C2} and \ref{C3} are considered. However, as the number of links to introduce between nodes in the initial graph $G$ depends on the number of nodes to introduce between them, and thus, on $h$ parity, the reduction is slightly modified.

$\bullet$ \textit{Definition of $V'$}\\
In this case, let $h'=h/2$, we first define $h'-1$ copies of $U$, denoted $U_i$ and $h'$ bijective functions $f_i$ with $i \in[1,h'-1]$:\\
		$\begin{array}{ccccc}
		f_i & : & V & \to & U_i\\
		& & v & \mapsto & f_i(v) = u_i\\
		\end{array}$\\
and the bijective function $f_{h'}$:\\
		$\begin{array}{ccccc}
		f_{h'} & : & E & \to & U_{h'}\\
		& & e & \mapsto & f_{h'}(e)= u_{h'}\\
		\end{array}$\\
We define the set $U_{h'}$. To each couple of nodes $(u_{h'-1},v_{h'-1})$ in $V_{h'-1}$, we associate a node $u_{h'}$ in $U_{h'}$ if and only if there is an edge between $f^{-1}_{h'-1}(u_{h'-1})$ and $f^{-1}_{h'-1}(v_{h'-1})$. Now, we can define the set $V_1 =  \{u_0\} \cup_i U_i$, $\forall i \in [1,h']$, where $u_0$ is a node introduced to model the data gathering tree in $G'$.\\

\vspace*{-0.4cm}
$\bullet$ \textit{Definition of $E'$}\\
To build the set $E'$, five types of links are introduced. We then have:\\
$E'= E_1 \cup E_2 \cup E_3 \cup E_4 \cup E_5$ where:\\
\begin{itemize}
\vspace*{-7pt}
\item $E_1 = \{(v,u_1)\ such\ that\ v \in V \ and\ u_1=f_1(v) \in U_1\}$. Thus, each node $v$ from the initial graph $G$ is linked to $u_{1}$, its associated node from the set $U_1$ (see links of type $e_1$ in Figure~\ref{6hopGraph}).
\item $E_2 = \cup_{l \in[1,h'-2]} \{(u_l,u_{l+1})\ such\ that\ u_l \in U_l\ and\ u_{l+1} \in U_{l+1}\ and\ f_l^{-1}(u_l)=f_{l+1}^{-1}(u_{l+1})\}$. Each node $u_{j}$ from $U_j$ is linked to node $u_{j+1}$ from $U_{j+1}$ associated with the same node $v \in V$, (see links of type $e_2$).
\item $E_3 = \{(u_{h'-1},u_{h'}), (u_{h'},v_{h'-1})\ such\ that\\\ u_{h'-1}\ and\ v_{h'-1} \in U_{h'-1}\ and\ u_{h'} \in U_{h'}\
with\ f_{h'}^{-1}(u_{h'})=(f_{h'-1}^{-1}(u_{h'-1}, f_{h'-1}^{-1}(v_{h'-1}))\}$. In other words, for each couple of nodes $u_{h'-1}$ and $v_{h'-1}$ in $U_{h'-1}$, we associate a node $u_{h'} \in U_{h'}$ if and only if $(f_{h'-1}^{-1}(u_{h'-1}), f_{h'-1}^{-1}(v_{h'-1})) \in E$. We then link $u_{h'}$ with $u_{h'-1}$ and $v_{h'-1}$ (see links of type $e_3$).
\item $E_4 = \{(u_i,u_j)\ such\ that\ u_i\ and\ u_j \in U_{h'}\ and \ i \ne j\}$. This means that the nodes in $U_{h'}$ form a complete graph (see links of type $e_4$).
\item $E_5 = \{(u_i,u_0) \forall \ u_i \in U_{h'}\}$. All nodes in $U_{h'}$ are linked to a node $u_0$ (see links of type $e_5$).
\end{itemize}
This construction is polynomial in time. The transformed graph $G'$ for $h=6$ of the initial graph $G$ depicted in  Figure~\ref{5hopGraph}.a is illustrated in Figure~\ref{6hopGraph}.

\vspace*{-8pt}

\vspace*{-15pt}
We now show, that the $k'$-color $h$-hop vertex coloring problem in both general and strategic modes, for any value of $h \ge 1$ has a solution if and only if the $k$-color 1-hop vertex coloring problem has a solution. We define the following Lemma:
\begin{lemma}
\label{lemmah-1}
All nodes in $G'\setminus G$ are at most $(h-1)$-hop neighbors.
\end{lemma}
\IEEEproof By construction of $G'$.
\endproof

\begin{lemma}
\label{lemmaG'}
To perform a $h$-hop coloring of the graph $G'$, the number of colors taken by nodes in $V_1$ is equal to $m$ with $m$ is equal to $(h'\cdot n)+1$ if $h$ is an odd number, and $(h'-1)\cdot n+p+1$ if $h$ is an even number, where $n$ is the number of nodes in $G$, $p$ is the number of edges in $G$.
\end{lemma}
\IEEEproof From Lemma~\ref{lemmah-1}, all nodes in $V_1$ are at most
$(h-1)$-hop neighbors. Hence, no color can be reused with
$h$-hop coloring ($h \ge 2$) of $G'$. By construction of $G'$, the number of these
nodes is equal to $(h'\cdot n)+1$ if $h$ is an odd number, and $(h'-1)\cdot n+p+1$ if $h$ is an even number.
\endproof

\begin{lemma}
\label{lemmaV'}
Any color used for a node in $V$ by a $h$-hop coloring of $G'=(V',E')$ cannot be used by any node in $V_1$.
\end{lemma}
\IEEEproof Let us consider any node $u \in  V' \setminus V$ and any node $v \in V$.
Let $d(v,u)$ be the number of hops between $v$ and $u$. By construction, $d(v,u)=d(v, f_1(v)) + d(f_1(v),u)$. From Lemma~\ref{lemmah-1}, $d(f_1(v),u) \leq h-1$ and since $f_1(v)$ is a neighbor of $v$, we get $d(u,v) \leq h$.
Hence, $u$ and $v$ must use different colors with $h$-hop coloring of $G'$ for $h \ge 1$.
\endproof

\noindent To complete the proof of Theorem~\ref{Thcomplexity}, we now prove the following Lemma:
\begin{lemma}
$G(V,E)$ has a one-hop
coloring with $k$ colors if and only if $G'(V',E')$ has a $h$-hop
coloring in general mode with $k'$ colors, with $h \ge 2$.
\end{lemma}
\IEEEproof Given a one-hop coloring of $G$ with $k$ colors, we want
to show that there exists a $h$-hop coloring of $G'$ with $k'$
colors as follows. According to Lemma~\ref{lemmaG'}, this $h$-hop coloring will use $k$ colors for
nodes in $V$ and $m$ colors for nodes in $
V'\setminus V$ with $m$ is equal to $(h'\cdot n)+1$ if $h$ is an odd number, and $(h'-1)\cdot n+p+1$ if $h$ is an even number. From Lemma~\ref{lemmaV'}, colors used in $V$ cannot be
reused in $V' \setminus V$. It follows that there exists a $h$-hop
coloring of $G'$ with exactly $k'=k+m$ colors.

Now, let us assume that we have a $h$-hop coloring of
$G'$ with $k'$ colors and we want to show that we can find a one-hop
coloring of $G$ with $k$ colors. From Lemma~\ref{lemmaG'}, $m$ colors are
needed for $h$-hop coloring of nodes in $V'\setminus V$. From
Lemma~\ref{lemmaV'}, colors used in $V$ cannot be reused in $V'\setminus V$.
Hence, $k'-m$ colors are used to color the nodes in $V$. Moreover, since
any two nodes $v_1$ and $v_2$ in $V$ that are one-hop neighbors in $G$
are $h$-hop neighbors in $G'$, by construction of $G'$, we deduce
that no two one-hop neighbors in $G$ use the same color. Hence, we
can find a valid one-hop coloring of $G$ with $k=k'- m$ colors.
\endproof

\begin{figure}[!h]
  \subfigure[]{\includegraphics[width=0.45\linewidth]{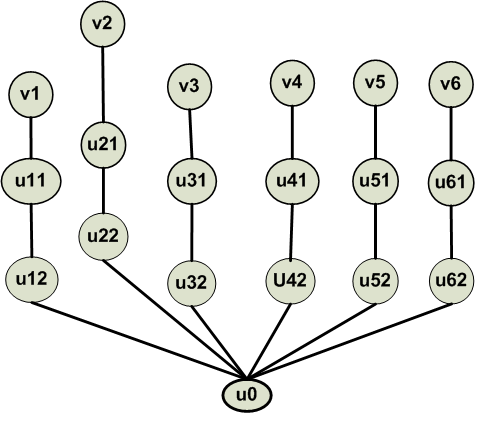}}
	\subfigure[]{\includegraphics[width=0.45\linewidth]{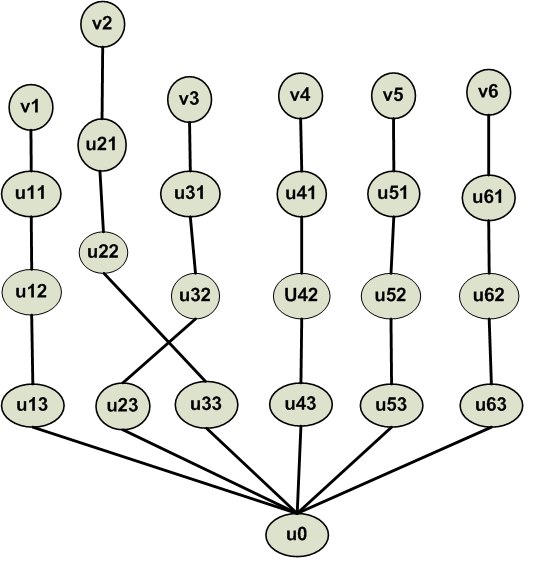}\label{6hop}}
	\vspace{-5pt}
	\caption{Tree built from $G'$: (a) $h$=5; (b) $h$=6.} \label{treeGraph}
\end{figure}

\begin{lemma}
$G(V,E)$ has a one-hop
coloring with $k$ colors if and only if $G'(V',E')$ has a $h$-hop
coloring in strategic mode with $k'$ colors, with $h \ge 1$.
\end{lemma}
\IEEEproof Given a one-hop coloring of $G$ with $k$ colors, we want
to show that there exists a $h$-hop coloring of $G'$ with $k'$
colors in strategic mode such that Constraint\ref{C0} is met, as follows.\\
We start by building a tree $T$ rooted at node $u_0$ from the graph $G'$. Nodes from $V$ are the leaves of this tree (see Figure~\ref{treeGraph}). In the case $h$ is odd, any node $v$ in $V$ has as parent $f_1(v)$. Any node $v_l$ in $U_l$ with $l$ a positive integer such that $1 \le l \le h'-1$ has as parent the associated node from the level $U_{l+1}$. Any node in $U_{h'}$ has as parent the root $u_0$.\\
In the case $h$ is even, $T$ has as root the node $u_0$, any node $v$ in $V$ has as parent $f_1(v)$, and any node in $U_l$ with $l$ a positive integer such that $1 \le l \le h'-2$ has as parent the associated node in $U_{l+1}$. Finally, we link the nodes in $U_{h'-1}$ to the tree. With any node $v_i \in U_{h'}$ ($1 \le i \le p$ where $p$ is the number of edges in $G$) linking two nodes $u_i$ and $u_j$ in $U_{h'-1}$ we associate as its child a node from the couple $(u_i,u_j)$, such that this node has not yet a parent.\\
To color $T$, we start by coloring the node $u_0$ the root of the tree. Then, we color nodes level by level, to finally reach the original nodes. From Lemma~\ref{lemmah-1} and Lemma~\ref{lemmaV'}, nodes in each level do not reuse colors from lower levels. Hence, each child has a color strictly higher than the color of its parent.
\endproof
Figure \ref{treeGraph} depicts the tree built from graph $G'$ for $h=5$ and $h=6$, where only tree links are represented.

\vspace*{-10pt}
\vspace*{-15pt}

\section{Theoretical results in grid topologies\label{Theoretical}}
Because of the NP-completeness of $h$-hop coloring in both general and strategic modes, proved in Section \ref{Complexity}, we focus on coloring algorithms based on heuristics. To compare different heuristics and select the best one, we use the optimal number of colors. The aim of this section is to determine this optimal number of colors for various topologies representative of wireless networks.
We start with regular topologies such as grids that constitute large or dense wireless networks. In a further work, we will extend our results to random topologies and study how to efficiently color a wireless network with a random dense topology by mapping it into a grid. For space reasons, we only consider $h$-hop coloring in general mode.

The goal of this section is to determine the optimal color number for the 3-hop coloring of grids with various transmission ranges.
In all the grids considered, we assume a transmission range higher than or equal to the grid step in order to get the radio connectivity. For simplicity reasons, the transmission range $R$ is expressed as a function of the grid step, that is considered as the unit. Hence, $R \geq 1$. Moreover, we assume an ideal environment where any node $u$ is able to communicate via a symmetric link to any node $v$ such that $d(u,v) \leq R$, where $d(u,v)$ denotes the euclidian distance from $u$ to $v$.

In this paper, we only study grid colorings that reproduce periodically a color pattern. As a consequence, the optimality of a coloring obtained is only true in the class of periodic colorings. We adopt the following notation and definitions:
\begin{definition}
\label{basicpatterndef}
A basic color pattern is the smallest color pattern that can be used to periodically tile the whole grid.
\end{definition}
\begin{definition}
\label{optimalpatterndef}
A basic color pattern is said optimal if and only if it generates an optimal periodic coloring of the grid.
\end{definition}

\subsection{Properties independent of the transmission range}
We can now give properties that do not depend on the transmission range value.
\begin{property}
Any color permutation of an optimal basic pattern is still valid and optimal.
\end{property}
\IEEEproof
With the color permutation, no two nodes that are 1-hop, 2-hop or 3-hop neighbors have the same color. Hence, the permuted coloring obtained is still valid. The permutation keeps unchanged the number of colors. Hence the coloring is still optimal.
\endproof
\begin{property}
\label{Pcolgrid}
Given an optimal color pattern of any grid and the color at node of coordinates (0,0), we can build a 3-hop coloring of a grid topology based on this pattern such that the color of node $(0,0)$ is the given color.
\end{property}
\IEEEproof
The coloring of the grid is obtained by setting the optimal color pattern in such a way that the color of node $(0,0)$ is the given color. The pattern is then reproduced to tile the whole topology.
\endproof
\begin{property}
\label{PropSingleRoundColoring}
Knowing an optimal color pattern of its grid and the color at node of coordinates $(0,0)$, each node can locally determine its own color based on its coordinates $(x,y)$. The 3-hop coloring obtained for the grid is optimal in terms of colors and rounds.
\end{property}
\IEEEproof The 3-hop coloring obtained for the grid only requires each node to know the color of node $(0,0)$, its coordinates in the grid and the optimal pattern to apply. Hence, it is optimal in terms of colors and rounds.
\endproof

\subsection{Optimal coloring for $R=1$, $R=1.5$ and $R=2$}
We now determine the optimal periodic coloring of grids for various transmission ranges. The proofs of the following theorems can be found in \cite{RR11}.
\begin{theorem}
The optimal 3-hop coloring of a grid topology with a transmission range $R=1$ requires exactly 8 colors, as shown in Figure~\ref{NeighborPatternFig}a. An optimal basic color pattern is given in Figure~\ref{NeighborPatternFig}b.
\begin{figure}[!h]
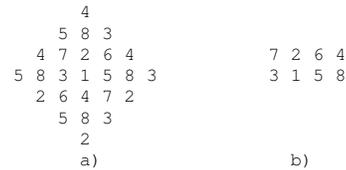

\vspace{-7pt}
\begin{scriptsize}
\begin{verbatim}
                  4
                5 8 3
              4 7 2 6 4            7 2 6 4
            5 8 3 1 5 8 3          3 1 5 8
              2 6 4 7 2
                5 8 3
                  2
                  a)                 b)
\end{verbatim}
\end{scriptsize}
\vspace{-5pt}
\caption{R=1 a) 3-hop coloring,  b) Basic color pattern with 8 colors.}
\label{NeighborPatternFig}
\end{figure}
\end{theorem}
\IEEEproof
Let $u$ be any non-border node of the grid. Let $\mathcal{N}(u)$ denote the set of nodes that cannot have the same color as $u$. The proof is done in four steps:
\begin{enumerate}
\item First step: At least 8 colors are needed to color node $u$ and $\mathcal{N}(u)$.
\item Second step: We build a valid coloring of $u$ and $\mathcal{N}(u)$ with 8 colors, as depicted in Figure~\ref{NeighborPatternFig}a.
\item Third step: This coloring can be regularly reproduced to constitute a valid coloring of the grid.
\item Fourth step: A basic color pattern containing exactly eight colors can be extracted (see Figure \ref{NeighborPatternFig}b).
\end{enumerate}
\vspace{-5pt}\endproof
\vspace{-5pt}\begin{theorem}
An optimal coloring of a grid with a transmission range $R=1.5$ needs exactly 16 colors, as shown in Figure~\ref{Color3hopR1.5Fig}a. An optimal basic color pattern is given in Figure~\ref{Color3hopR1.5Fig}b.

\begin{figure}[!h]
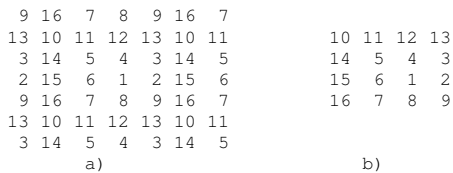

\vspace{-7pt}
\begin{scriptsize}
\begin{verbatim}
             9 16  7  8  9 16  7
            13 10 11 12 13 10 11         10 11 12 13
             3 14  5  4  3 14  5         14  5  4  3
             2 15  6  1  2 15  6         15  6  1  2
             9 16  7  8  9 16  7         16  7  8  9
            13 10 11 12 13 10 11
             3 14  5  4  3 14  5
                   a)                       b)
\end{verbatim}
\end{scriptsize}
\vspace{-5pt}
\caption{R=1.5  a) 3-hop coloring,  b) Basic color pattern with 16 colors.}
\label{Color3hopR1.5Fig}
\end{figure}
\end{theorem}
\vspace{-10pt}
\begin{theorem}
An optimal coloring of a grid with a transmission range $R=2$ needs exactly 25 colors, as shown in Figure~\ref{Color3hopR2Fig}a. An optimal basic color pattern is given in Figure~\ref{Color3hopR2Fig}b.

\begin{figure}[!h]
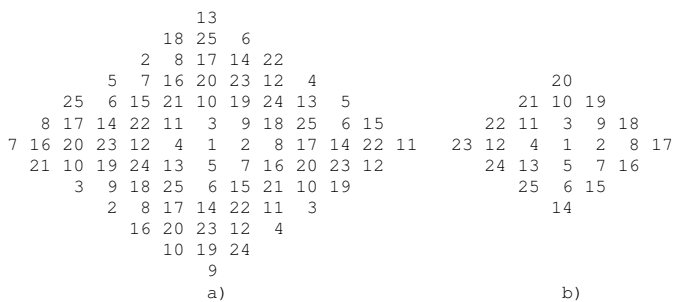

\vspace{-7pt}
\begin{scriptsize}
\begin{verbatim}
                  13
               18 25  6
             2  8 17 14 22
          5  7 16 20 23 12  4                     20
      25  6 15 21 10 19 24 13  5               21 10 19
    8 17 14 22 11  3  9 18 25  6 15         22 11  3  9 18
 7 16 20 23 12  4  1  2  8 17 14 22 11   23 12  4  1  2  8 17
   21 10 19 24 13  5  7 16 20 23 12         24 13  5  7 16
       3  9 18 25  6 15 21 10 19               25  6 15
          2  8 17 14 22 11  3                     14
            16 20 23 12  4
               10 19 24
                   9
                   a)                              b)
\end{verbatim}
\end{scriptsize}
\vspace{-5pt}
\caption{R=2  a) 3-hop coloring,  b) Basic color pattern with 25 colors.}
\label{Color3hopR2Fig}
\end{figure}
\end{theorem}

\section{SERENA: a three-hop node coloring algorithm\label{SERENA}}
We now focus on wireless ad hoc and sensor networks, where the algorithm complexity must be kept small. Because of the NP-completeness of $h$-hop coloring, heuristics are used to color network nodes. SERENA is a distributed 3-hop node coloring based on an heuristic: the nodes with the highest priority are colored first. Results in Section \ref{Theoretical} giving the optimal number of colors allow us to compare different node priority assignments and to select the best one for SERENA: the one giving a number of colors close to the optimal.

\subsection{Basic principles of 3-hop node coloring}
As previously said, 3-hop node coloring is necessary to support unicast transmissions with immediate acknowledgement in case of general communications, where any node is likely to exchange information with any neighbor node.
In SERENA, any node $u$ proceeds as follows to color itself:
\begin{enumerate}
\item Node $u$ characterizes the set $\mathcal{N}(u)$ of nodes that cannot have the same color as itself. This set depends on the type of:
\begin{itemize}
\item \textit{communications supported}: unicast and/or broadcast;
\item \textit{application}: general where any node is likely to exchange information with any neighbor node or on the contrary tree type where a node exchanges information only with its parent and its children in the data gathering tree;
\item \textit{acknowledgement for unicast transmissions}: immediate or deferred.
\end{itemize}
In our case, the set $\mathcal{N}(u)$ is the set of neighbors up to 3-hop.
\item Node $u$ computes its priority. Different priority assignments will be tested in the next subsection.
\item Node $u$ applies the two following rules:
\begin{itemize}
\item \textbf{Rule R1}: Node $u$ colors itself if and only if it has a priority strictly higher than any uncolored node in $\mathcal{N}(u)$.
\item \textbf{Rule R2}: To color itself, node $u$ takes the smallest color unused in $\mathcal{N}(u)$.
\end{itemize}
\end{enumerate}
In SERENA, each node sends its Color message to its 1-hop neighbors. This message contains the information related to the priority and color of the node itself and of its 1-hop and 2-hop neighbors.

\subsection{Performance evaluation in grid topologies}

Performance of SERENA is evaluated by the number of colors and the number of rounds needed to color all network nodes. The optimal number of colors is given for comparison. Table~\ref{TabSerenaGrid} reports the results obtained by SERENA for different grids and different transmission ranges. The number of nodes ranges from 100 to 900. The density is computed as the average number of neighbors per node (i.e. average number of nodes in radio range of a sender). It varies from 3 to 25. $Prio1$ means that there is only one criterion used for node priority assignment: either the position of the node in the grid line, or a random number. These choices are justified by the node address assignment: addresses are assigned either consecutively according to grid lines or randomly. $Prio2$ means that two criteria are used: first the number of nodes up to 2-hop and second either the position of the node in the grid line or a random number. Notice that identical results have been obtained when considering the column instead of the line in the grid. Results are averaged over 10 simulation runs. We can draw two conclusions from the results obtained:\\
$a)$ The number of colors strongly depends on the density of nodes and weakly on the number of nodes. The number of rounds strongly depends on the number of nodes and weakly on the density. For instance, for a radio range=1, we get 58 rounds for a 10x10 grid with 100 nodes and 178 rounds for a 30x30 grid with 900 nodes.\\ 
$b)$ The optimal number of colors is obtained by SERENA for $Prio2+line$ as well as $Prio1+line$ for transmission ranges $R=1$ and $R=1.5$.\\
$c)$ The coloring method that colors the whole grid in a single round by repeating the basic color pattern (see Property \ref{PropSingleRoundColoring}) uses exactly the same number of colors whatever the grid size. However, in small topologies where the number of nodes is too small to enable the repetition of the basic color pattern in the four directions, SERENA with $Prio2+random$ is able to obtain a number of colors smaller than this method. This is explained by the fact that a color can be reused earlier: a conflicting node does not exist. This phenomenon can be observed in the simulation results for a grid size 10x10, for $R=2.5$ or $3$.\\
$d)$ The number of colors as well as the number of rounds depend on the priority assignment. Another address assignment produces another coloring. The average number of colors reaches 15.4 for a 30x30 grid for the random priority assignment while it is 8 when the priority is given by the position of the node in the line. Why do not we have the same number of colors? The reason is given by the node priority assignment. A pure random assignment $Prio1+random$ gives the worst results, except in configurations described in $c)$. The line assignment gives better results than the random one, suggesting that regularity of the topology must be taken into account. $Prio2$ outperforms $Prio1$ in 12 cases out of 15 both in terms of colors and rounds, suggesting that nodes from the center, having the highest number of constraints must be colored first, as found in \cite{hans04} for 1-hop coloring. Such nodes will then enforce their colors on the border nodes.

\vspace*{-7pt}
\begin{table}[!ht]
\caption{Number of colors and rounds obtained by SERENA for various grids and transmission ranges.}
\label{TabSerenaGrid}
\vspace*{-15pt}
\begin{center}
\begin{tabular}{|c|c|c|c|c|c|c|c|}
\hline
Radio&Grid& Dens.&Optim &\multicolumn{2}{|c|}{Prio1}& \multicolumn{2}{|c|}{Prio2}\\
\cline{5-8}
range&size& & & line& random & line&random\\
\hline
1 &10x10 & 3.6& 8C & 8C & 13.8C & 8C& 11.8C \\
        &    & &   &58R & 67.4R & 54R& 77.8R \\
   \cline{2-8}
   & 20x20& 3.8&8C & 8C& 15.4C & 8C & 14.8C \\
         &  &  &   &118R & 82.4R& 114R& 110R\\
   \cline{2-8}
   & 30x30&3.87 &8C &8C& 15.4C& 8C & 15.4C \\
         &   & &   &178R& 93R& 174R& 116.6R\\
\hline
1.5 &10x10 & 6.84&16C &16C & 23.6C & 16C & 19.4C\\
          &     &  &  &67R & 94.8R& 65R & 107.6R\\
   \cline{2-8}
   & 20x20& 7.41&16C &16C& 27.6C & 16C & 26.6C\\
         &   &   & &137R& 144.6R& 134R & 166.6R\\
   \cline{2-8}
   & 30x30& 7.6&16C& 16C& 28.4C &16C& 27.8C\\
         & &  &   &207R& 169.2R& 204R & 195.4R\\
\hline
2 &10x10 & 10.04&25C &30C& 33.8C &30C &28.2C \\
        &    &   &   &85R& 136.4R&123R&121.8R\\
   \cline{2-8}
   & 20x20 & 11.01&25C & 33C& 41.8C&33C&36.4C\\
          &    &  &   & 175R& 236.6R&174R&282.4R \\
   \cline{2-8}
   & 30x30& 11.34&25C & 33C& 44.4C&34C&42.8C\\
         &     &  &  & 265R& 278R&264R&376.6R\\
\hline
2.5 &10x10& 15.8 & 45C &52C& 50.8C &49C &42.33C \\
        &       & &    &94R& 176.8R&105R&146.17R\\
   \cline{2-8}
   & 20x20 & 17.85& 45C & 54C& 66.8C&54C&64.8C\\
          &     &   &  & 194R& 348.2R&197R&359.8R \\
   \cline{2-8}
   & 30x30& 18.56& 45C & 55C& 74C&58C&73.6C\\
         &      & &    & 294R& 422.2R&297R&440.8R\\
\hline
3 &10x10 & 21.16&68C &70C& 68C &71C &67.6C \\
        &     & &    &112R& 193R&116R&197.8R\\
   \cline{2-8}
   & 20x20 & 24.49&68C & 80C& 93.4C&79C&93C\\
          &  &  &     & 232R& 449.4R&231R&451.4R \\
   \cline{2-8}
   & 30x30&25.64&68C & 83C& 107.8C&81C&107.8C\\
         &    &  &   & 352R&601.2R&351R&601.4R\\
\hline
\end{tabular}
\end{center}
\end{table}

\vspace*{-15pt}
\section{Conclusion\label{Conclusion}}
In this paper we have proved that the $h$-hop node coloring problem in both general and strategic modes is NP-complete, for any integer $h>0$. We have then focused on specific cases of large or dense wireless networks: grids with a radio range higher than the grid step. We have determined an optimal periodic 3-hop coloring of grids with various transmission ranges. We have then compared the results obtained by SERENA, a distributed 3-hop coloring algorithm for different node priority assignments. The priority assignment equal to the number of neighbors up to 2-hop, where ties are broken by addresses assigned by grid line, gives the optimal number of colors for a transmission range of 1 or 1.5. It also outperforms random priority assignments in 12 cases over the 15 cases tested. As a further work, we will optimize SERENA to take into account these results. We will also study how to map a grid on a given random topology and determine the best grid adapted to this topology.

\end{document}